# USING BIOPHOTONICS TO STUDY SIGNALING MECHANISMS IN A SINGLE LIVING CELL

Donald C. Chang

*Department of Biology, Hong Kong University of Science and Technology,*

*Clear Water Bay, Hong Kong, China*

To illustrate the power of the biophysical approach in solving important problems in life science, I present here one of our current research projects as an example. We have developed special biophotonic techniques to study the dynamic properties of signaling proteins in a single living cell. Such a study allowed us to gain new insight into the signaling mechanism that regulates programmed cell death.

## 1. Introduction

The basic unit of life is a cell. The function of the cell depends on a very complicated system of signals, which consist of proteins, RNA, DNA as well as other chemical compounds including sugars and lipids. The most important objective in life science research today is to understand how this cellular signaling system works. At present, most of the obtained information is based on biochemical studies that are most suited to characterize isolated components (such as genes and proteins) extracted from cells. Since the cell is a highly organized system in which the interaction between signaling molecules are spatially dependent, the true signaling system cannot be fully understood without preserving the structural integrity of the living cell. Thus, a new challenge in bio-medical research now is to extend the traditional *in vitro* biochemical studies into *in vivo* biophysical studies, so that we can examine the mechanisms of signal transduction in the intact living cell. Only with such an *in vivo* approach can we measure the dynamic changes of the signaling molecules within the sub-cellular environment in response to a specific signaling event.

In recent years, our laboratory has been engaging in developing innovative optical techniques to study the molecular signaling systems in a single living cell. These techniques include labeling specific proteins with GFP (green fluorescence protein) using gene-fusion, and measurements of protein-protein interaction using the FRET (fluorescence resonance energy transfer) method. Our approach has certain advantages. First, since these optical methods are non-invasive, we can preserve the organization and structure of the cell during the biological study. Thus, the information obtained is highly reliable. Second, by conducting the study in a single cell, we can avoid the problem of cell synchrony, which would be difficult to achieve in a cell population study. Third, in the single cell study, we can correlate the temporal- and spatial-dependent changes of a specific signaling molecule with a particular cellular event. Finally, using the FRET technique, we can measure the dynamics of a specific protein-protein interaction or the activation of a given enzyme within a single cell. Such measurements are not possible using conventional biochemical methods. To demonstrate the usefulness of this biophotonic approach, in the following, we will present the results of our study on the signaling mechanism of programmed cell death as an example.

## 2. The signaling mechanism of programmed cell death

One of the hottest problems in life science today is to understand the control mechanisms of programmed cell death (also called "apoptosis"). Programmed cell death is a controlled cell suicidal process that has great importance in maintaining the normal physiological function of a human being. It allows the biological organism to destroy damaged or unwanted cells in an orderly way. For example, apoptosis is used in the thymus to eliminate self-reactive T cells to avoid auto-immunity [1]. Furthermore, apoptosis plays a critical role in animal development; it is utilized to eliminate the unwanted cells so that the proper structure of functional organs can be formed in the embryo. For example, formation of digits in a hand and the proper targeting of the neural connection are known to be dependent on apoptosis. Thus, defects in apoptosis will result in major developmental abnormalities [2].

The study of apoptosis has gained widespread attention in recent years due to its newly discovered roles in a variety of pathological processes. For example, when DNA is damaged in a cell and cannot be repaired, the cell will enter apoptosis to avoid the formation of abnormalities in the tissue. Thus, failure of programmed cell death can cause cancer. Besides tumor genesis, malfunction of apoptosis is also found to be associated with numerous pathological disorders, including Alzheimer's disease, auto-immune disease, AIDS, etc.

In the last few years, a large number of studies have been conducted aiming to understand the process of apoptosis at a molecular basis. The signaling pathways that direct the programmed cell death process turn out to be very complicated. There are many external signals that can trigger the initiation of apoptosis, including UV-irradiation, activation of the "death domain" via the TNF (tumor necrosis factor) receptor, treatment of hormone (e.g. glucocorticoid) and chemotherapy drugs (e.g. camptothecin) [3, 4] (see Figure 1). As for internal signals, we know that apoptosis is the outcome of a programmed cascade of intracellular events, which are centered on the activation of a class of cysteine proteases called "caspases" [5]. Some of these caspases (such as caspase-8 and caspase-9) are "initiators" of the apoptotic process, while others (such as caspase-3) are "executioners". Besides caspases, a number of gene products are also known to be key players in processing apoptosis. These include the Bcl-2/Bax family proteins, p53, IAP (Inhibitor of Apoptosis Protein) and Smac/Diablo [6-8]. When these genes are over-expressed (or mutated), some of them can cause cells to undergo apoptosis, while others can prevent cells from entering programmed cell death. Some of the organelles are also known to play an important role in the progression of apoptosis [9]. Among them, mitochondria plays the most important role in inducing apoptosis, since some of the substances discharged from mitochondria are key activators of the downstream apoptotic pathways [9, 10]. A summary of the major signaling pathway of apoptosis is shown in Figure 1. At present, the detailed molecular mechanisms of apoptosis regulated by various internal and external signals are still under active investigation.

In this paper, we summarized a series of studies to demonstrate how we use optical technologies for studying signaling mechanisms in programmed cell death. These studies were focused on revealing the mechanism through which mitochondria release certain apoptotic protein factors during UV-induced apoptosis. Mitochondria were known to play a central role in the apoptotic process. During programmed cell death, the releases of two mitochondrial intermembrane proteins, cytochrome c and Smac from the mitochondria into the cytosol are the key steps that trigger the activation of a series of caspase cascades, which commits the cell to the death process (Figure 1). Once cytochrome c is in the cytosol, it will bind to Apaf-1. In the presence of dATP, this complex recruits procaspase 9 and activates it. Active caspase 9 then proteolytically activates execution caspases including caspase 3, which finally causes cell death. As to Smac, it can facilitate activation of caspases indirectly, by neutralizing a set of caspase inhibitor, known as IAP. At present, it is known that the release of mitochondrial proteins is regulated by the Bcl-2/Bax family proteins. The mechanism, however, is not yet clear. There is also a question on whether different types of apoptotic proteins were released from mitochondria through the same pathway. In this study, we have gained important insight in these problems by examining the dynamic re-distribution of GFP-labeled Bax and Smac during apoptosis using living cell-imaging techniques.

## 3. Using the GFP-gene fusion technique to monitor the dynamic redistribution of signaling proteins in a single living cell

Green fluorescent protein (GFP) is a naturally fluorescent protein isolated from the jellyfish *Aequorea victoria* [11]. In the last 10 years, GFP has been commonly used as a fluorescent marker in molecular biology, cell biology and medicine. This is mainly due to the fact that the chromophore of GFP is formed by an internal posttranslational autocatalytic cyclization, which does not require any exogenous substrates or enzymes [12]. Fused with other proteins, GFP and its variants can be used to monitor the dynamic processes of signaling proteins in many cell types and organisms.

Besides the automatically formation of chromophore, several other properties make GFP an excellent reporter molecule. One of the great advantages of GFP is its high stability. GFP fluorescence is very stable in the presence of denaturants and proteases, as well as over a range of pH and temperatures. GFP is also highly tolerant to some classical fixatives such as formaldehyde and glutaraldehyde, and fluorescence quenching agents, which provides an opportunity to view and localize GFP in preserved tissue. Furthermore, the GFP chromophore is relatively photostable during imaging. This high stability of GFP is correlated with its unique three-dimensional structure, as shown in Figure 2A. Eleven strands of β-sheet form an antiparallel barrel with short helices forming lids on each end [13]. The chromophore, formed by cyclization of 65Ser-Tyr-Gly67, is near geometric enter of the cylinder. The tightly constructed β barrel appears to protect the chromophore well, providing overall stability. Another advantage of GFP is that no cofactors or exogenously added substrate are required for detecting GFP fluorescence, which is perfect for live-cell assays by exposing cells to minimal invasive treatment. Moreover, the GFP reporter, despite its relatively large size, usually does not interfere with the normal activity or mobility of the tagged proteins and can be added to either the C- or N-terminus of target proteins.

Wild-type GFP (wtGFP) absorbs ultraviolet (UV) and blue light with a maximum peak of absorbance at 395 nm and a minor peak at 470 nm, and emits green light at 509 nm with a shoulder at 540 nm. Recently, random and site-directed mutageneses have produced useful GFP mutants that are brighter and have excitation and emission spectra different from wtGFP (see Figure 2B). Particularly useful are the enhanced GFP color variants: EYFP (yellow), EGFP (green), ECFP (cyan), and EBFP (blue). Improved through mutagenesis, these proteins are some of the most widely used reporters in biological research. They have been optimized for brighter emission, faster chromophore maturation and more resistance to photobleaching. In addition to GFP and its variants, there are many other fluorescent proteins. DsRed, a brilliantly red fluorescent protein cloned from *Discosoma coral*, for example, represents a new marker that can be used together with GFP variants for multicolor imaging.

In mammalian cells, GFP technology has been used in a wide variety of applications, including studying the subcellular distribution, function, and expression of proteins, time-lapse imaging for studying the dynamic redistribution of cytoplasmic, cytoskeletal or organelle proteins, double- or triple-labeling to compare the distribution and dynamics of different proteins simultaneously within cells, and measuring protein-protein interactions through FRET.

4. **Using biophotonics to study the mechanism of mitochondrial permeabilization during apoptosis: Testing of the "swelling model"**

As discussed earlier, mitochondria are known to play a central role in apoptosis. However, what regulate mitochondria to release cytochrome c and Smac and cause irreversible cellular damage is still not well understood. It was only observed that the outer membrane of mitochondria became permeabilized during apoptosis so that some of the intermembrane proteins (including cytochrome c and Smac) were released into the cytosol. According to the literature, several models of apoptosis-induced MOM (mitochondrial outer membrane) permeabilization had been proposed [9]. They are:

(a) Model of matrix swelling, i.e., apoptotic stress may induce mitochondrial matrix swelling which in turns causes nonspecific rupture of the MOM;
(b) A member of the Bcl-2 family, Bax, may insert itself into the outer mitochondrial membrane, resulting in forming a Bax-oligomer channel;
(c) Bax may bind with VDAC (voltage-dependent anion channel) to form a hybrid channel;
(d) Bax, or Bax-like family proteins, may aggregate to the MOM and destabilize it by forming a "lipid-protein complex"

The basis for model (a) was that, a megachannel traversing across the outer and inner mitochondrial membranes called "PT (permeability transition) pore" is induced to open during apoptosis [15]. Solutes and water in the cytosol then enter mitochondrial matrix through this channel, causing the mitochondrion to swell and rupture its outer membrane [15,16]. As a result, cytochrome c and other proteins in the inter-membrane space are released. This hypothesis has been widely cited and debated in the literature [15-24], but so far, there is still a lack of conclusive evidence to prove or disprove it. First, the supporting evidence for the mitochondrial swelling theory was based mainly on indirect observations. For example, specific inhibitors of the PT pore such as cyclosporin A or bongkrekic acid were found to prevent apoptosis in hepatocytes and other cell models [25,26], while PT pore-opening agent like attractyloside or $Ca^{2+}$ was observed to induce matrix swelling and apoptosis [17,26]. Conflicting data, however, were also reported and suggested that cytochrome c release may not be related to PT pore opening. For example, Eskes et al [23] observed that Bax-induced cytochrome c release cannot be inhibited by either cyclosporin A or bongkrekic acid in isolated mitochondria.

Second, there was a controversy on whether the mitochondria indeed swelled during apoptosis or not. Some studies report the observation of mitochondrial swelling [16,17], while others reported that swelling never occurred [20,24]. Third, most previous studies were done using fixed cells or isolated mitochondria [16-20, 23, 24, 27], which could become swelled during the fixation or isolation processes [28]. Forth, even if mitochondria were observed to swell during apoptosis, it was still not clear whether the mitochondrial swelling was the cause or the result of cytochrome c release.

In order to overcome these difficulties and directly test the mitochondrial swelling theory, one must study the events of cytochrome c release and mitochondrial swelling in an intact living cell. Thus, we used photonic techniques to measure the dynamic re-distribution of GFP-labeled cytochrome c during UV-induced apoptosis in living HeLa cells, and used a red color fluorescent dye, Mitotracker, to monitor the morphological change of mitochondria at the same time. The objective of our study is to examine two important questions: (1) Do mitochondria swell during apoptosis? (2) If they do, is the mitochondrial swelling the cause or consequence of cytochrome c release?

The GFP-tagged cytochrome c protein (Cyt c-GFP) was produced by expressing a plasmid vector containing the cytochrome c-GFP fusion gene in HeLa cells. We exposed the cells to UV light (300 $\mu W/cm^2$) for 3 minutes to induce apoptosis and then recorded the distribution of Cyt c-GFP at different times using a confocal microscope. Figure 2C shows a sample record of this time series measurement. Before cytochrome c was released from mitochondria, it appeared as filamentous structures. As cytochrome c was released from mitochondria, the Cyt c-GFP became uniformly distributed over the entire cytoplasm. This releasing process was found to finish in 5 minutes.

We then determined whether mitochondrial matrix became swollen during apoptosis, and if yes, whether the swelling of mitochondria occurred before or after the release of cytochrome c. We used a laser scanning confocal microscope (MRC 600, Bio-rad) to simultaneously measure the dynamic re-distribution of Cyt c-GFP and morphological change of mitochondria monitored by mitotracker staining within living HeLa cells during UV-induced apoptosis [14]. Figure 2D showed that the morphology of mitochondria still remained filamentous and mitochondria did not swell when cytochrome c was released from mitochondria to cytosol during apoptosis. This finding suggests that mitochondrial swelling could not be responsible for the release of cytochrome c.

To further examine the temporal relationship between cytochrome c release and mitochondrial swelling, we conducted a quantitative analysis of the time-dependent change of the distribution of cytochrome c and the morphological change of mitochondria within a single living cell. The method is demonstrated in Fig. 3. First, we determined the diameter of the mitochondria from the magnified Mitotracker images using a line-scan method (Fig. 3A-C). The diameter of a mitochondrion was taken as the half width of its pixel profile based on the line-scan measurement (Fig. 3C). Several line-scan measurements were made on a single mitochondrion to obtain its average diameter. This procedure was repeated on many mitochondria within the same cell to obtain a statistical value. Second, to quantify the distribution of cytochrome c, we examined the pixel distribution profiles of the cyt c-GFP image in two sample regions of the cell: Region 1 contained only cytoplasm, while Region 2 contained both cytoplasm and mitochondria (Fig. 3D). By comparing the pixel distribution profiles of these two regions, we can choose a proper threshold to separate the fluorescence signal contributed by cyt c-GFP located in the mitochondria from that of the cytosolic cyt c-GFP.

We conducted this quantitative analysis for a stack of images recorded at different time on a single HeLa cell undergoing UV-induced apoptosis. This analysis allows us to have an accurate determination of the temporal relationship between cytochrome c release and mitochondrial swelling at a single-organelle level. In fact, one could measure both the cytochrome c distribution and morphological change in a single mitochondrion. For example, the mitochondrion in the upper right corner of Figs. 2D was shown to release its cytochrome c first without undergoing a morphological change. Results of a quantitative analysis of its cytochrome c release and change in diameter as a function of time were summarized in Fig. 3E. It is clear evident that mitochondrial swelling occurred after cytochrome c was released. This living cell measurement thus strongly indicates that cytochrome c release in apoptosis was not caused by a PT pore opening-triggered swelling of the mitochondrial matrix.

**5. Using biophotonics to study the mechanism of mitochondrial permeabilization during apoptosis: Testing of the "channel model"**

Next, we tested whether the formation of specific channels or pores by Bax-like family proteins in the MOM is responsible for the release of cytochrome c and Smac. To determine whether Bax formed channel in the MOM, we asked two specific questions:

*(i) How large is the molecular complex formed by Bax?*
*(ii) How large is the "pore"?*

To answer the first question, we examined the dynamic distribution pattern of GFP-labeled Bax during UV-induced apoptosis within a single living cell. Figure 4A shows a sample record of this time series measurement. Initially, Bax was diffusely distributed in the cytosol. At a later time, some GFP-Bax proteins began to translocate from the cytosol to mitochondria and formed small aggregates. Soon after that, more Bax aggregated in mitochondria to form large clusters. Within minutes, the cell shrank and died.

After detecting the dynamic re-distribution of GFP-Bax from cytosol to mitochondria, we investigated which stage of Bax localization was responsible for the release of cytochrome c or Smac during UV-induced apoptosis. By labeling the different proteins with different fluorescent probes, including CFP, GFP, YFP and RFP (a red fluorescent protein), we compared the temporal relationship between Bax translocation and the release of cytochrome c and Smac from mitochondria during UV-induced apoptosis. We found that the release of cytochrome c occurred at the same time as the formation of the small Bax aggregates (data not shown). Furthermore, by simultaneously measuring the re-distribution of GFP-Bax and the intensity change of the fluorescent dye Mitotracker red (Fig. 4B), we found that the formation of small Bax aggregates in the mitochondria was coinciding with the decrease of the mitochondrial membrane potential. These findings suggest that the small Bax aggregates formed in the mitochondria is responsible for the permeabilization of the MOM.

Then, we used an imaging method to determine the size of this Bax aggregate. By comparing the size of the Bax aggregate with man-made fluorescent beads of known diameters (supplied by Molecular Probes), we concluded that the size of the Bax aggregate was about 0.2 – 0.3 micrometers. Then, using droplets of purified GFP solution of known protein concentration, we determined the number of Bax molecules in the Bax aggregates by comparing their fluorescent intensities with that of the droplets. We found that each Bax aggregate contained about 100 to 400 molecules. These findings suggest that the Bax aggregate responsible for the release of cytochrome c was much larger than a channel formed by Bax oligomer.

To answer the second question (i.e., How large is the pore?) We examined what is the size of the molecule that can pass through the "pore". Using different color mutants of GFP, we can label two different signal proteins and measure their dynamic redistributions simultaneously within a single living cell (Fig. 4C & 4D). Using this method, we determined the kinetics of Bax translocation and the release of Smac from mitochondria during UV-induced apoptosis. We found that the formation of Bax aggregates in mitochondria was simultaneously associated with the release of Smac from mitochondria to the cytosol (Figs. 4B, 4E & 4F). In fact, we found that the pathway of releasing Smac is essentially the same as that of releasing cytochrome c [29]. In view of these findings, we think that models (b) and (c) may be discarded. These two models assume that the mitochondrial proteins were released through a "channel" formed by oligomer of Bax or Bax associated with VDAC. It is conceivable that such channel may be large enough to pass a small molecule such as cytochrome c (MW=15 kDa). But, the channel would be very difficult to allow Smac to pass through. It is known that the endogenous Smac generally appeared as tetramers in the cellular environment, its molecular weight is about 100 kDa [8]. For the YFP-labeled Smac, it would have even a larger size (~200 kDa). From crystal structure studies, it was suggested that Smac may form dimers instead of tetramers [30]. But even if the dimer model is true in the *in vivo* environment, the molecular weight of the YFP- Smac dimer would still approach 100kDa. It is impossible for such a large protein complex to pass through a channel. Hence, we believe only the model (d) (i.e., Bax forming a "lipid-protein complex") could fit with our findings. Recently, Kuwana et al. reported that Bax could produce mega openings at the mitochondrial outer membrane, which could release supramolecules up to 2000 kDa [31]. With such a model, one can expect that Smac and Cytochrome c are released together upon the aggregation of Bax in the outer mitochondrial membrane.

**Acknowledgements**

I thank Drs. Wenhua Gao, Liying Zhou and Lingli Zhou for their assistance in the experimental work, and Mr. Wenbin Cao for his help in preparing the manuscript. This work was supported by the Research Grants Council of Hong Kong (HKUST6466/05M, N_HKUST616/05) and the EHIA project of HKUST.

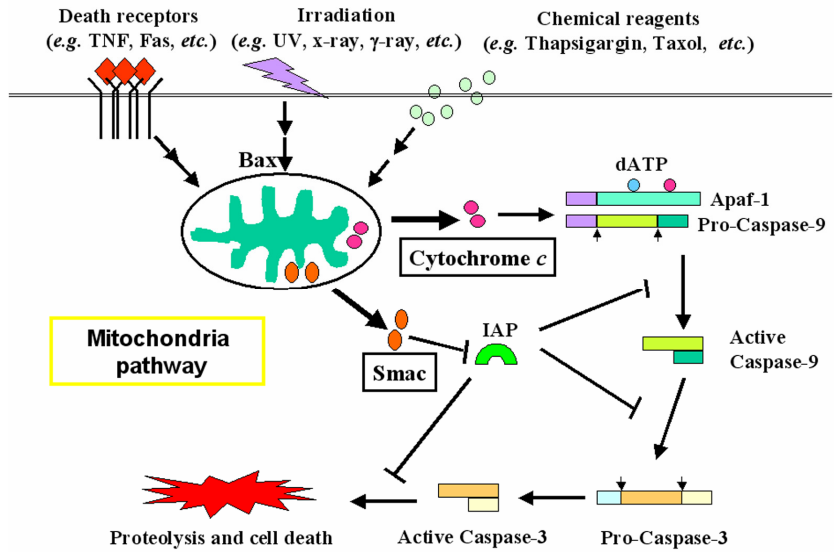

**Figure 1.** A schematic diagram showing the mitochondria-dependent signaling pathway in programmed cell death.

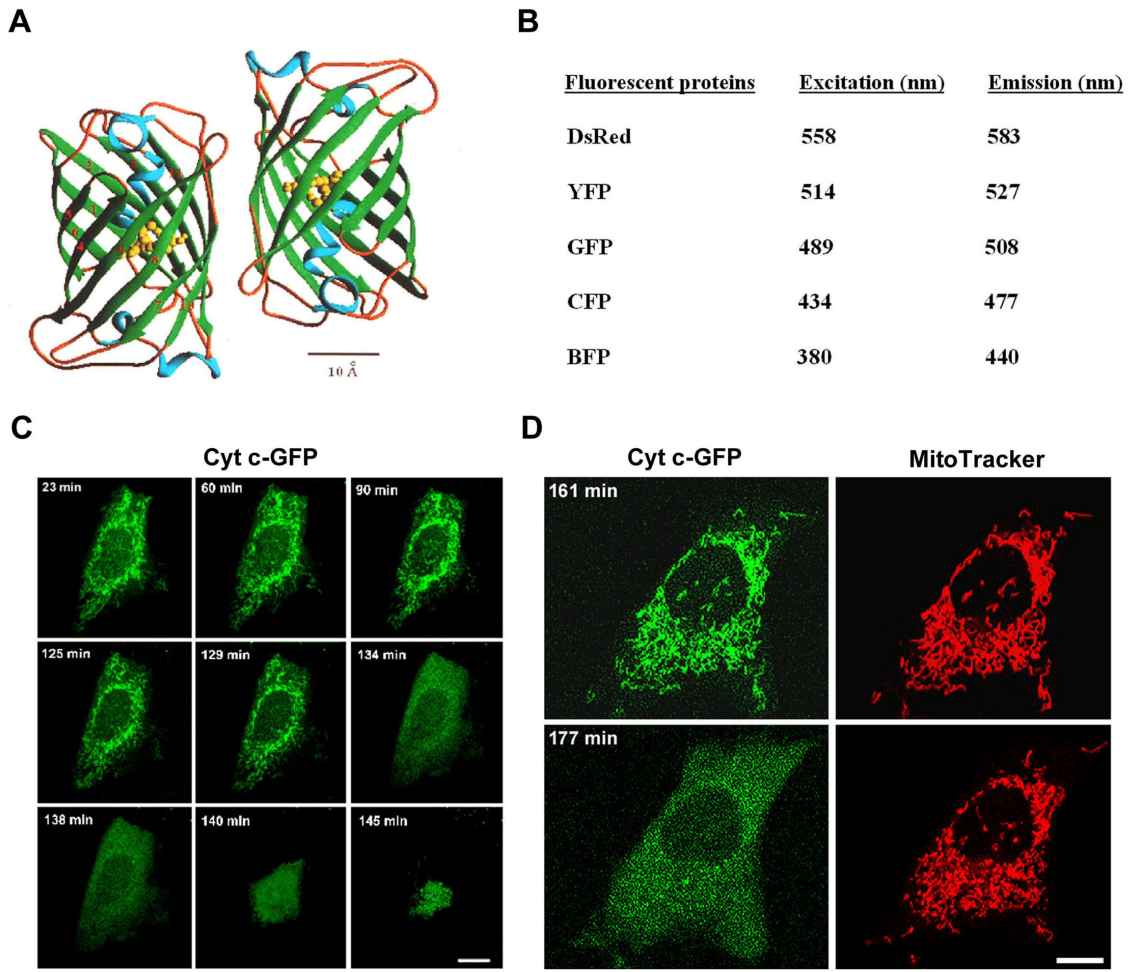

**Figure 2.** (**A**) The three-dimensional structure of GFP. [From Yang et al., 1996]. Reproduced with permission from Nature Biotechnology. (**B**) Fluorescence proteins which have different excitation and emission wavelengths. (**C**) Sample records of a time-lapse confocal measurement of Cyt c-GFP distribution in a single living HeLa cell during UV-induced apoptosis. (**D**) Sample records of a time-lapse confocal measurement of both Cyt C-GFP distribution (left) and Mitotracker distribution (right) in a living HeLa cell during UV-induced apoptosis. Elapsed time following the UV treatment is shown in each panel. Scale bar, 20 μm.

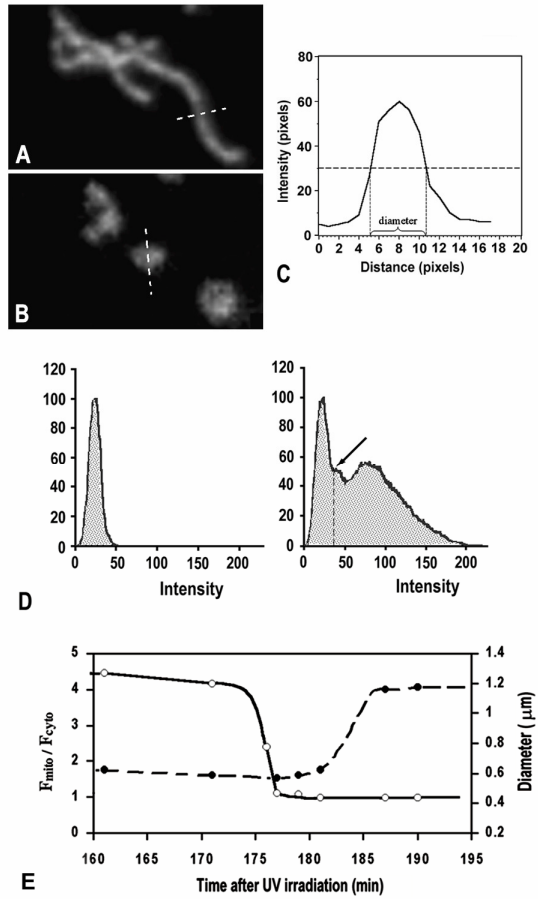

**Figure 3.** Quantitative analysis of the change of mitochondrial diameter in response to cytoc-GFP release. (**A & B**) Magnified images of Mitotracker before (**A**) and after (**B**) cytoc-GFP release from mitochondria. (**C**) The diameter of a mitochondrion was taken as the half width of its pixel profile based on a line-scan measurement of the Mitotracker image. (**D**) The distribution profiles of cytoc-GFP fluorescence intensity in a region containing only cytoplasm (left), or in a region containing both cytoplasm and mitochondria (right). The ordinate represents the relative population of pixels. The arrow marks the threshold that separates the fluorescence signal contributed by cytoc-GFP located in the mitochondria ($F_{mito}$) from that of the cytosolic cytoc-GFP ($F_{cyto}$). (**E**) Time-dependent changes of cytoc-GFP distribution and mitochondrial diameter as measured from a single mitochondrion.

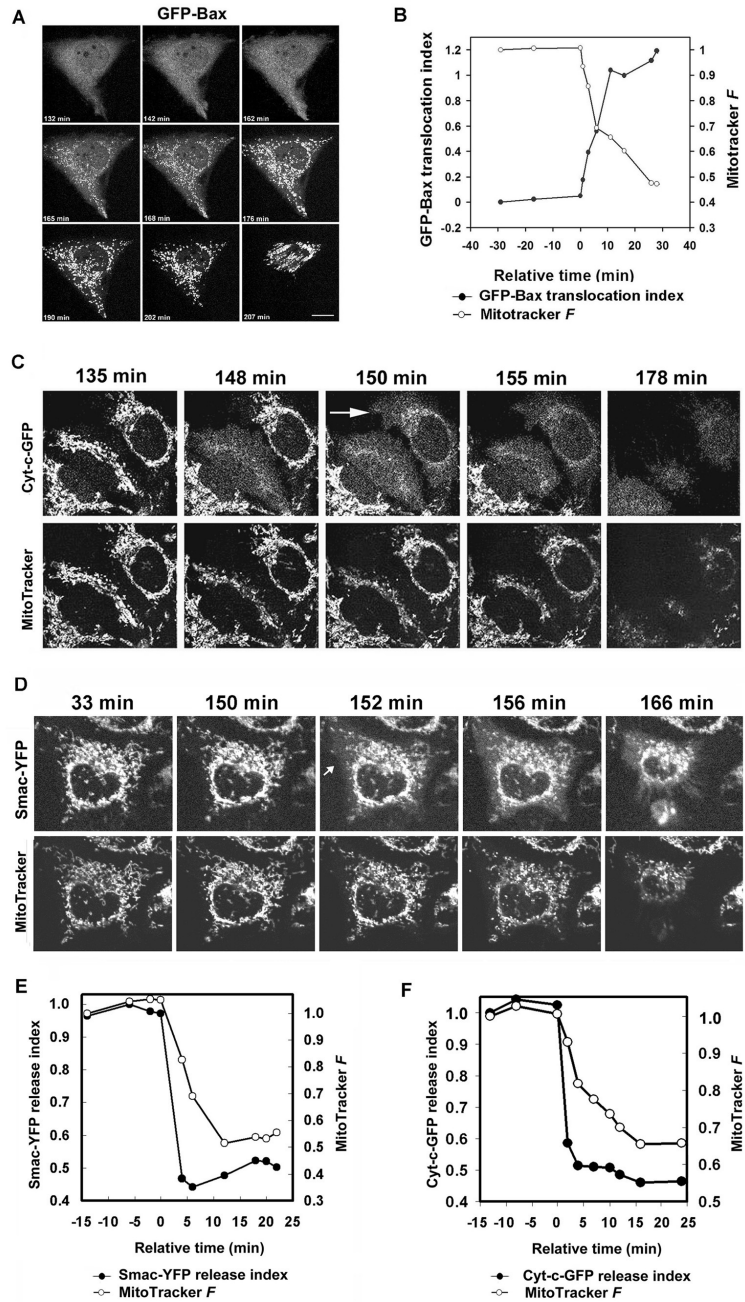

**Figure 4.** (**A**) Sample records of confocal measurements of GFP-Bax distribution in a single living HeLa cell following UV-treatment. Scale bar, 10 μm. (**B**) Temporal relationship between Bax aggregation and the reduction of the mitochondrial membrane potential (as indicated by Mitotracker fluorescence). (**C**) Time lapse measurement of Cyt c-GFP re-distribution and Mitotracker signal in living HeLa cells. (**D**) Time lapse confocal measurement of Smac-YFP re-distribution in living HeLa cells. (**E,F**) Temporal relationship between Smac-YFP (or Cyt c-GFP) release and the reduction of Mitotracker signal within a living HeLa cell during UV-induced apoptosis.